\documentclass[letterpaper,14pt]{article}
\usepackage{setspace}
\doublespace
\usepackage[superscript]{cite}
\usepackage{amsmath}
\usepackage{amsfonts}
\usepackage{amssymb,multirow,wrapfig}
\usepackage{color}
\usepackage{graphicx}
\usepackage[small,normal,up]{caption2}
\usepackage{cancel}
\usepackage{adjustbox}
\usepackage{array}
\usepackage{booktabs}
\usepackage{multirow}

\newcommand{\bra}[1]{\ensuremath{\left< #1 \right|}}
\newcommand{\ket}[1]{\ensuremath{\left| #1 \right>}}

\newcommand{\dd}{\dagger}

\newcommand{\be}[0]{\begin{equation}}
\newcommand{\ee}[0]{\end{equation}}
\newcommand{\bea}[0]{\begin{eqnarray}}
\newcommand{\eea}[0]{\end{eqnarray}}

\newcommand{\degree}{^{\circ}}
\newcommand{\expval}[1]{\left\langle #1 \right\rangle}
\newcommand{\sig}[3]{\sigma_{\text{#1}}^{\text{#2},\text{#3}}}

\begin{document}

\begin{center}

\LARGE\textbf{Multi-photon entanglement in high dimensions}

\vspace{4mm} \normalsize\textbf{Mehul Malik,$^{1,2\ast}$ Manuel Erhard,$^{1,2}$ Marcus Huber,$^{3,4}$\\Mario Krenn,$^{1,2}$ Robert Fickler,$^{1,2,\dd}$ Anton Zeilinger$^{1,2}$}\\

\vspace{1mm}
\footnotesize
\textit{$^1$Institute for Quantum Optics and Quantum Information (IQOQI), Austrian Academy of Sciences, Boltzmanngasse 3, A-1090 Vienna, Austria}\\
\textit{$^{2}$Faculty of Physics, University of Vienna, Boltzmanngasse 5, 1090 Vienna, Austria}\\
\textit{$^{3}$Universitat Autonoma de Barcelona, 08193 Bellaterra, Barcelona, Spain}\\
\textit{$^{4}$ICFO-Institut de Ciencies Fotoniques, 08860 Castelldefels, Barcelona, Spain}\\
\textit{$^*$mehul.malik@univie.ac.at}\\
\textit{$^\dd$Present address: Department of Physics and Max Planck Centre for Extreme and Quantum Photonics, University of Ottawa, Ottawa, K1N 6N5, Canada}

\end{center}
\vspace{4mm}

\vspace{8mm}

\newpage
\noindent \textbf{
Entanglement lies at the heart of quantum mechanics---as a fundamental tool for testing its deep rift with classical physics\cite{Einstein:1935hx,Bell:1964wu}, while also providing a key resource for quantum technologies such as quantum computation\cite{Bennett:2000kl} and cryptography\cite{Ekert:1991kl}. In 1987 Greenberger, Horne, and Zeilinger\cite{Greenberger:1989vx} realized that the entanglement of more than two particles implies a
non-statistical conflict between local realism and quantum mechanics. The resulting predictions were experimentally confirmed by entangling three photons in their polarization\cite{Pan:2000do}. Experimental efforts since have singularly focused on increasing the number of particles entangled, while remaining in a two-dimensional space for each particle \cite{Kelly:2015gi, Yao:2012fp, Lanyon:2014eh}. Here we show the experimental generation of the first multi-photon entangled state where both---the number of particles and the number of dimensions---are greater than two. Interestingly, our state exhibits an asymmetric entanglement structure that is only possible when one considers multi-particle entangled states in high dimensions\cite{Huber:2013ie}. Two photons in our state reside in a three-dimensional space, while the third lives in two dimensions. Our method relies on combining two pairs of photons, high-dimensionally entangled in their orbital angular momentum, in such a way that information about their origin is erased\cite{Zeilinger:1997eb}. Additionally, we show how this state enables a new type of ``layered" quantum cryptographic protocol where two parties share an additional layer of secure information over that already shared by all three parties\cite{Hillery:1999cb}. In addition to their application in novel quantum communication protocols, such asymmetric entangled states serve as a manifestation of the complex dance of correlations that can exist within quantum mechanics.
} 


In the recent past, big mysteries in quantum mechanics have been illuminated by taking small steps in the right direction. The phenomenon of quantum interference, for example, appeared when one considered a single quantum particle with at least two discrete levels, or dimensions. Moving to two particles gave us quantum entanglement\cite{Einstein:1935hx} and Bell's inequalities\cite{Bell:1964wu}, which allowed the conflict between quantum mechanics and local realism to be tested in a statistical manner\cite{Clauser:1969ff}. Increasing the number of entangled particles to three, while seemingly a simple step, provided the first ``all-or-nothing" test of local realism\cite{Pan:2000do}. Alongside this, increasing the dimensions of a single quantum particle from two to three provided the first clear test of quantum contextuality\cite{Lapkiewicz:2011iq}. History would dictate that increasing both the number of quantum particles and the number of dimensions in concert will lead to further interesting and fundamental quantum phenomena. In this letter, we discuss one such phenomenon that arises when we consider three-particle entangled states of dimension greater than two---namely, asymmetric multi-particle entanglement.

Two-dimensional entangled states have been studied in great detail due to their natural application in quantum information and computation. However, the amount of information carried by a photon is potentially enormous, and the ability to harness this information leads to quantum communication systems with record capacities and unprecedented levels of security\cite{Mirhosseini:2015fy}. A natural space for exploring large dimensions in a quantum system is a photon's spatial degree of freedom\cite{Malik:2014ht}. The orbital angular momentum (OAM) of a photon is a spatial property that provides a discrete and unbounded state space\cite{MolinaTerriza:2007ig}. The phase of an OAM-carrying photon winds azimuthally from $0$ to $2\pi\ell$ around the axis. The number of ``twists" in the wavefront are given by the quantum number $\ell$ and dictate the photon state dimension. Recent experiments have shown the entanglement of two photons in up to $100\times100$ dimensions in their spatial modes\cite{Dada:2011dn,Krenn:2014jy}. The dimensionality of two-photon entangled states is given by a single number, the Schmidt number\cite{Terhal:2000gd}, which is the rank of the reduced single particle density matrix. This number represents the minimum number of levels one needs to faithfully represent the state and its correlations in any local basis. When one considers three entangled photons of dimensionality greater than two, the question of how many levels per photon are involved has three answers\cite{Huber:2013ie}. In our experiment, we aim to create a state which has the form
\be\label{332i} \ket{\Psi}_{332} = \frac{1}{\sqrt{3}}\big[\ket{0}_A\ket{0}_B\ket{0}_C+\ket{1}_A\ket{1}_B\ket{1}_C+\ket{2}_A\ket{2}_B\ket{1}_C\big]. \ee
Notice that the first two photons, $A$ and $B$, live in a three-dimensional space, while the third photon, $C$, lives in a two-dimensional space. The dimensionality of this state is given by a vector of three numbers $(3,3,2)$. Specifically, these are the ordered ranks of the single particle reductions of the state density operator:
\be \textrm{rank}(\rho_{A})=3,\;\;\;\;\;\;\textrm{rank}(\rho_{B})=3,\;\;\;\;\;\;\textrm{rank}(\rho_{C})=2,\ee
where $\rho_{i} = \text{Tr}_{\bar{i}}\ket{\Psi}_{332}\bra{\Psi}_{332}$ is the state of system $i\in(A,B,C)$. Only certain combinations of these three ranks are allowed, leading to a rich structure of possible asymmetric high-dimensional multipartite entangled states\cite{Cadney:2014iw}. In addition to being of fundamental interest, such states enable a novel ``layered" quantum communication protocol that allows the sharing of secure information asymmetrically amongst multiple parties\cite{LMQKD}. We discuss this protocol in the context of our state later in the text.

The entanglement of three photons was first achieved by combining two pairs of polarization-entangled photons in such a way that it became impossible, even in principle, to know which pair one of the detected photons belonged to\cite{Zeilinger:1997eb,Bouwmeester:1999jq}. The workhorse of such experiments is the polarizing beam-splitter (PBS), which was designed to separate a light beam into its horizontal and vertical polarization components. Interestingly, a PBS can also be used to mix the polarizations of two input photons from independent polarization-entangled pairs in such a manner that an output photon contains polarization components from both input photons. In this manner, information about their origin is erased, producing a four-photon, two-dimensional GHZ state\cite{Pan:2001vk}. The largest such state created thus far used seven PBSs to entangle eight photons in their polarization\cite{Yao:2012fp}. In order to manipulate the high-dimensional space of OAM, one would need a device akin to the PBS, but operating on a photon's spatial wavefunction as opposed to its polarization. 

\begin{figure}[t!]
\centering\includegraphics[scale=.65]{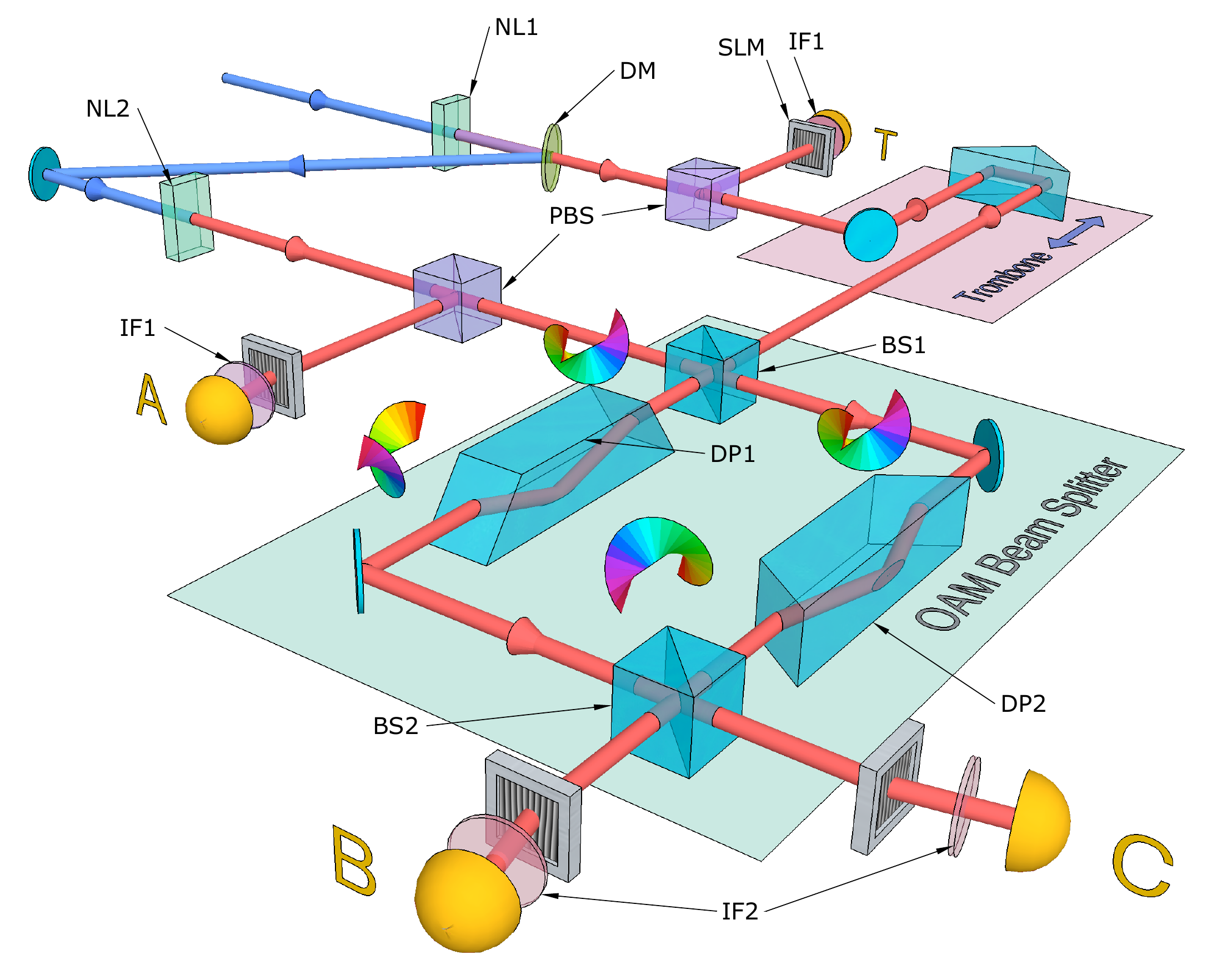}
\caption{\textbf{Schematic of the experiment}. A blue laser pulse creates two pairs of OAM-entangled photons in two nonlinear crystals (NL1 and NL2). Both pairs are spatially separated via polarizing beam splitters (PBS). A moving trombone prism ensures that a photon from each pair arrives simultaneously at the OAM beam splitter. This device contains two dove-prisms (DP1 and DP2). DP1 reflects an incoming photon that inverts its spiral phase front, while DP2 both inverts and rotates it by $180^\circ$. This results in interference that depends on the parity (odd or even) of the spatial mode. Considered as a whole, the OAM beam splitter reflects odd components of OAM and transmits even ones, mixing the OAM components of two input photons. A coincidence at detectors B and C can only arise when both detected photons have the same mode parity. This projects photons at detectors A, B, and C into an asymmetric three-photon entangled state that is entangled in $3\times3\times2$ dimensions of its OAM. The detection of a trigger photon at detector T heralds the presence of this state (see extended data figure 1 for an in-depth schematic).}\label{setup}
\end{figure}  

In our experiment we choose to use a simple Mach-Zehnder interferometer that was designed to sort photons based on the parity of their OAM quantum number\cite{Leach:2002wy}. As shown in Fig.~\ref{setup}, this interferometer has dove prisms (DP1 and DP2) in both arms, rotated by $90^\circ$ with respect to one another. DP1 reflects an incoming photon that inverts its spiral phase front, while DP2 both inverts and rotates it by $180^\circ$. Thus, each input photon interferes with a rotated version of itself, leading to constructive or destructive interference based on whether its spatial mode has odd or even parity. We use this device as a two-input, two-output ``OAM beam splitter" that reflects photons with odd OAM values and transmits even ones. In this manner, an outgoing photon contains a superposition of odd and even values of OAM from two independent input photons.

Our procedure for generating the high-dimensional tripartite state uses two independent entangled photon pairs created in two non-linear crystals (NL1 and NL2). Suppose that the two pairs are in the state
\bea\label{tensor9} \ket{\Psi}_{ABCD}=&\frac{1}{\sqrt{3}} \big( \ket{1}_A\ket{\text{-}1}_B +\ket{0}_A\ket{0}_B +\ket{\text{-}1}_A\ket{1}_B \big)\nonumber\\
&\otimes\frac{1}{\sqrt{3}} \big(\ket{1}_C\ket{\text{-}1}_D +\ket{0}_C\ket{0}_D +\ket{\text{-}1}_C\ket{1}_D \big),
\eea
which is a tensor product of two OAM-entangled photons pairs $(A, B$ and $C,D)$ of dimension $d=3$ each (OAM quantum number $\ell\in\{-1,0,1\}$). At this stage, mixing these two states results in nine possible four-photon amplitudes. Photons $B$ and $C$ are then incident on the two inputs of the OAM beam splitter. Since this device reflects odd values of OAM and transmits even ones, a coincidence detection between the two outputs can only arise when photons $B$ and $C$ are both carrying either odd or even values of OAM. This projects the state from Eq.~\eqref{tensor9} onto a subspace spanned by the five terms $(\ket{1}_A\ket{\text{-}1}_B\ket{1}_C\ket{\text{-}1}_D)$, $(\ket{1}_A\ket{\text{-}1}_B\ket{\text{-}1}_C\ket{1}_D)$, $(\ket{0}_A\ket{0}_B\ket{0}_C\ket{0}_D)$, $(\ket{\text{-}1}_A\ket{1}_B\ket{1}_C\ket{\text{-}1}_D)$, and $(\ket{\text{-}1}_A\ket{1}_B\ket{\text{-}1}_C\ket{1}_D)$. Photon $D$ is then measured in a superposition state given by $\ket{P}^{0,\text{-}1}_D=\frac{1}{\sqrt{2}}(\ket{0}_D+\ket{\text{-}1}_D)$ via a mode-projection carried out by a spatial light modulator (SLM) and a single-photon detector (T). This acts as a trigger for the three-photon entangled state:

\begin{figure}[b!]
\centering\includegraphics[scale=1]{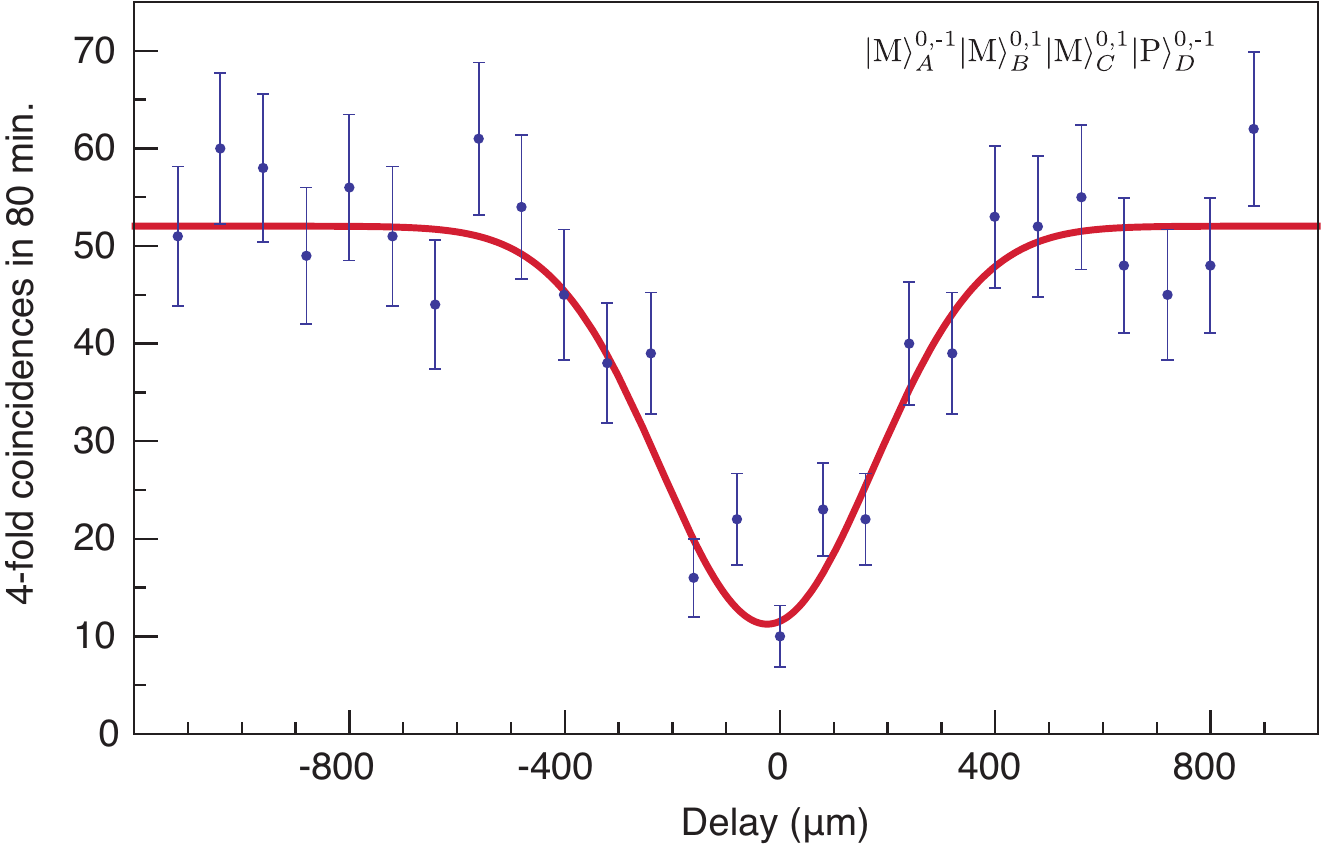}
\caption{\textbf{Three-photon coherent superposition.} Experimental data showing the result of measuring photons $A$, $B$, and $C$ in states $\ket{\text{M}}^{0,\text{-}1}_A$, $\ket{\text{M}}_B^{0,1}$, and $\ket{\text{M}}_C^{0,1}$ respectively, conditioned on measuring photon $D$ in state $\ket{\text{P}}_D^{0,\text{-}1}$. The drop in coincidence counts at the position of zero delay results from interference between photons $B$ and $C$. This indicates that the 3-photon state from equation~\ref{332} is in a coherent superposition. A Gaussian fit is calculated based on the spectral shape of our narrowband filters and has a full-width at half-maximum of 473$\mu$m and a visibility of 63.5\% (see methods section for details).}\label{HOM}
\end{figure}

\be\label{332} \ket{\Psi}_\text{exp} = \frac{1}{\sqrt{3}}\big[\ket{1}_A\ket{\text{-}1}_B\ket{1}_C+\ket{0}_A\ket{0}_B\ket{0}_C+\ket{\text{-}1}_A\ket{1}_B\ket{1}_C\big]. \ee

Note that this state has the same form as equation~\ref{332i}, except photons $A$ and $B$ are in a three-dimensional space given by the OAM quantum numbers $(\text{-}1,0,1)$, while photon $C$ is in a two-dimensional space given by $(0,1)$. In order to ensure that these three terms are in a coherent superposition (as opposed to an incoherent mixture), we perform superposition measurements in a two-dimensional subspace spanned by the second and third terms in equation~\ref{332}. Measuring photon $A$ in a superposition state given by $\ket{M}^{0,\text{-}1}_A=\frac{1}{\sqrt{2}}(\ket{0}_A-\ket{\text{-}1}_A)$ projects the three-photon state into:

\be\label{PM} \frac{1}{\sqrt{3}}\ket{\text{M}}^{0,\text{-}1}_A(\ket{\text{P}}^{0,1}_B\ket{\text{M}}^{0,1}_C+\ket{\text{M}}^{0,1}_B\ket{\text{P}}^{0,1}_C), \ee

\begin{figure}[b!]
\centering\includegraphics[scale=.8]{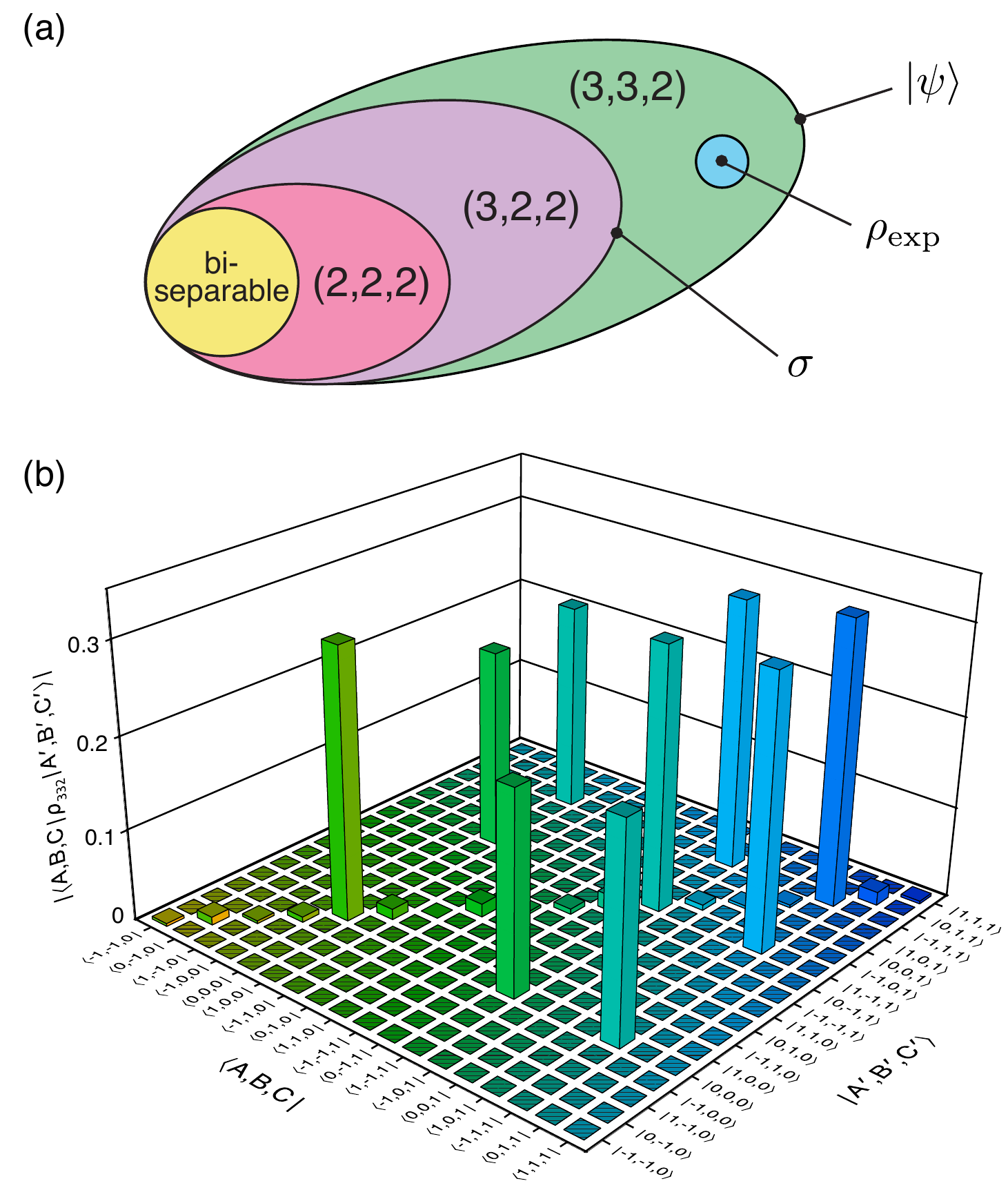}
\caption{\textbf{Witnessing genuine multipartite entanglement in high dimensions.} (a) In order to verify that our three-photon state is entangled in $3\times3\times2$ dimensions, we have to show that it cannot be decomposed into entangled states of a smaller dimensionality structure. First, we calculate the best achievable overlap of a $(322)$-state $\sigma$ with an ideal target $(332)$-state $\ket{\Psi}$ to be $F_\text{max}=2/3$. Next, we calculate the overlap $F_\text{exp}$ of our experimentally generated state $\rho_{\text{exp}}$ with the target state $\ket{\Psi}$. (b) The 18 diagonal and 3 unique off-diagonal elements of $\rho_\text{exp}$ that are measured in order to calculate a value of $F_\text{exp}=0.801\pm0.018$ (elements not measured are filled in with lines). This is above the bound of $F_\text{max}=0.667$ by 7 standard deviations and verifies that the generated state is genuinely multipartite entanglement in $3\times3\times2$ dimensions of its orbital angular momentum. $F_\text{exp}$ does not reach its maximal value of 1 because the state superstition is not perfectly coherent.}\label{matrix}
\end{figure}


\noindent where $\ket{P}^{0,1}_{B/C}=\frac{1}{\sqrt{2}}(\ket{0}_{B/C}+\ket{1}_{B/C})$ and $\ket{M}^{0,1}_{B/C}=\frac{1}{\sqrt{2}}(\ket{0}_{B/C}-\ket{1}_{B/C})$. The terms $\ket{\text{P}}^{0,1}_B\ket{\text{P}}^{0,1}_C$ and $\ket{\text{M}}^{0,1}_B\ket{\text{M}}^{0,1}_C$ are missing because of two-photon destructive interference at the OAM beam splitter, which only occurs when photons $B$ and $C$ are indistinguishable\cite{Hong:1987gm}. Note that this requires both two-photon and one-photon interference (for two different photons) to occur at the OAM beam splitter, which proved to be an experimental challenge. The use of narrowband interference filters (IF2) before detectors B and C blurs out the temporal correlations of the photons with their entangled partners to a certain extent, ensuring that they cannot be distinguished based on their time of arrival\cite{Kaltenbaek:2009vf}. We look for destructive interference between photons $B$ and $C$ by projecting them into states $\ket{\text{M}}^{0,1}_B$ and $\ket{\text{M}}^{0,1}_C$, and scanning the trombone prism shown in Fig.~\ref{setup}. At the zero-delay position, both photons arrive at the first beam splitter (BS1) at precisely the same time and interfere destructively, leading to a drop in coincidence counts between detectors B and C, conditioned on photons $A$ and $D$ being detected in states $\ket{\text{M}}^{0,\text{-}1}_A$ and $\ket{\text{P}}^{0,\text{-}1}_D$ respectively. The characteristic dip in Fig.~\ref{HOM} has a width of $473\mu\text{m}$ that matches what we expect from the bandwidth of our filters and the thickness of our crystals (see Methods section for details).


In order to verify that our state is indeed genuinely multipartite entangled in a $3\times3\times2$ dimensional space, we have to prove that it cannot be decomposed into entangled states of a smaller dimensionality structure\cite{Huber:2013ie, Fickler:2014eq}, such as a $(322)$-state or a two-dimensional GHZ/$(222)$-state. As shown in Fig.~\ref{matrix}(a), these states take on a ``russian doll" structure of concentric subsets embedded in the set of (332)-states, and can be ruled out as follows. First, we find the best achievable overlap $F_\text{max}$ of a $(322)$-state $\sigma$ with an ideal $(332)$-state $\ket{\Psi}$. In the Methods section, we show that $F_\text{max}=2/3$. Next, we calculate the overlap $F_\text{exp}$ of our experimentally generated state $\rho_{\text{exp}}=\ket{\Psi}_{\text{exp}}\bra{\Psi}_{\text{exp}}$ with an ideal $(332)$-state. If $F_\text{exp}>F_\text{max}$, our state is certified to be entangled in $3\times3\times2$ dimensions. In order to calculate $F_\text{exp}$, it is sufficient to measure the 18 diagonal elements of $\rho_\text{exp}$, as well as its 3 unique off-diagonal elements that are expected to be non-zero. The 18 diagonal elements are measured directly through projective measurements of 27 minutes each. The 3 off-diagonal elements are measured via a series of 144 projective measurements (see Methods section for details). These elements are plotted in Fig.~\ref{matrix}(b), and numerical values are provided in extended data table 1. From these, we calculate an experimental fidelity $F_\text{exp}=0.801\pm0.018$, which is above the bound of $F_\text{max}=0.667$ by 7 standard deviations. This certifies that our three-photon state is indeed entangled in $3\times3\times2$ dimensions of its orbital angular momentum.

\begin{figure}[t!]
\centering\includegraphics[scale=.4]{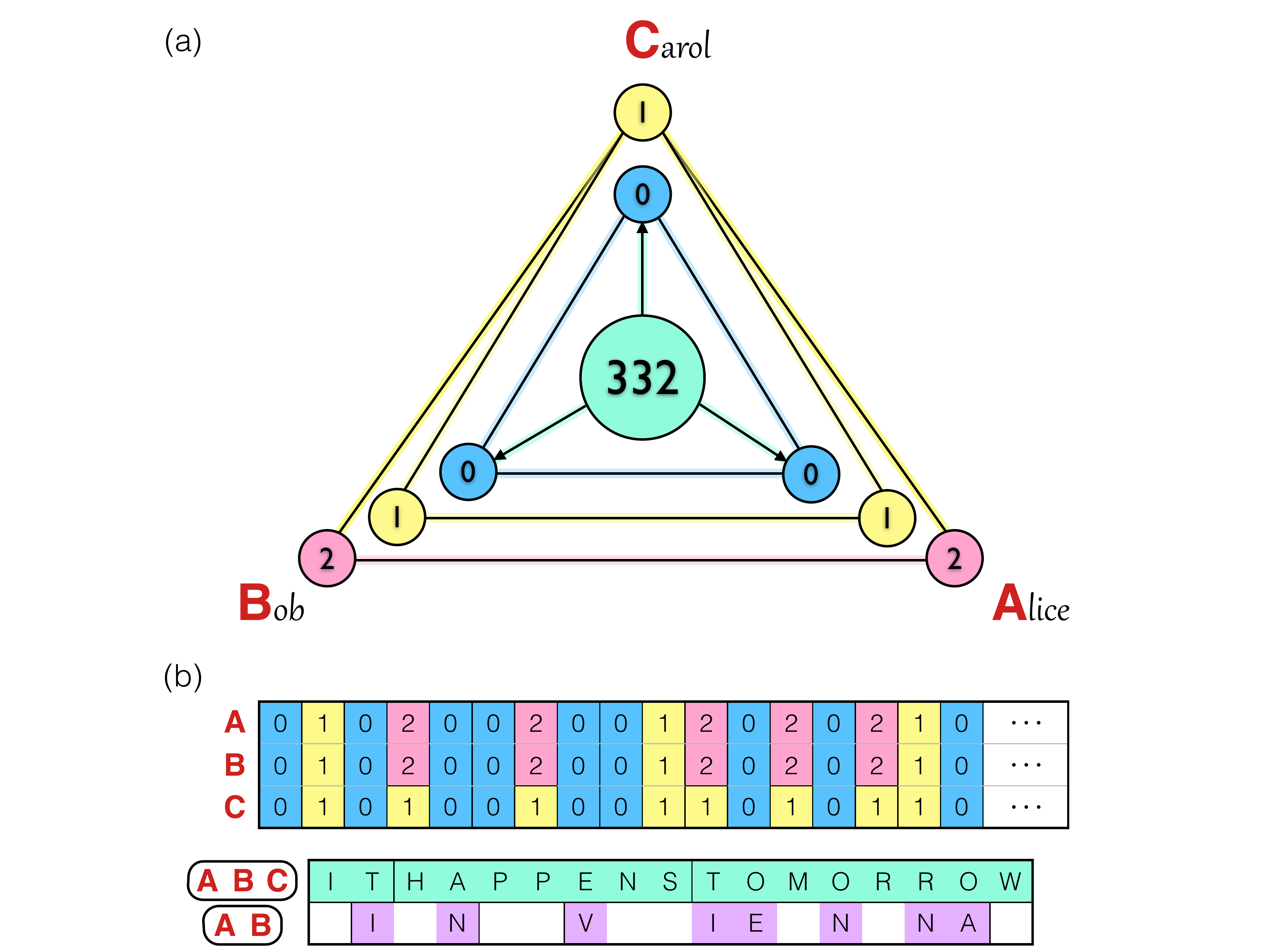}
\caption{\textbf{A layered quantum communication protocol.} (a) When Alice, Bob, and Carol share an asymmetric $(3,3,2)$ entangled state, all three have access to an entangled bit in the $(0,1)$ basis. Additionally, whenever Carol has a 1, Alice and Bob share an entangled bit in the $(1,2)$ basis. (b) After sufficient data is collected, these bits can be used as a one-time pad to encrypt a secret message shared among all three parties. In addition, Alice and Bob can share a second layer of information unknown to Carol. Security is verified by checking for the presence of $(3,3,2)$ entanglement in a randomly selected subset of photons.}\label{LMQKD}
\end{figure}  

Finally, this high-dimensional three-photon entangled state enables a new type of ``layered" quantum cryptographic protocol where layers of information can be shared asymmetrically with unconditional security\cite{LMQKD}. Consider the communication scheme depicted in Fig.~\ref{LMQKD}. If Alice, Bob, and Carol share a $(332)$-entangled state, all three parties have one entangled bit in the $(0,1)$ basis. However, Alice and Bob have access to an additional entangled bit in the $(1,2)$ basis whenever Carol has a 1. This state allows all three parties to generate a secure string of random bits that can be used as a key. Two-thirds of the time, however, Alice and Bob are able to generate a second secure key. Thus, all three parties can securely share a secret message, with Alice and Bob sharing an additional layer of information that is unknown to Carol. Security can be tested by verifying the presence of $(332)$-entanglement in a randomly selected subset of photons using the entanglement witness from above. The implementation of such a protocol would require multi-outcome measurements of OAM, techniques for which are quite mature today \cite{Mirhosseini:2013em}.

We have created a three-photon entangled state and verified that it is high-dimensionally genuinely multipartite entangled in its orbital angular momentum. This is the first such entangled state where both, the number of particles and the number of dimensions, are greater than two. In addition, our generated state displays a unique asymmetry in its entanglement structure that has not been experimentally observed before. This asymmetric entanglement has enabled us to develop a new layered quantum communication protocol where different layers of information can be shared securely amongst multiple parties. Our experimental method can be modified to create a vast array of even more complex entangled states\cite{Melvin}---of which the state above is just one example. Such asymmetric entangled states constitute a new direction in experimental studies of entanglement, and will allow the development of complex, multi-level quantum networks in the future.


\section*{Acknowledgements}

We thank T.~Scheidl, M.~Tillman, R.~Lapkiewicz, and G.B.~Lemos for helpful discussions. MM acknowledges funding from the European Commission through a Marie Curie fellowship (OAMGHZ). MH acknowledges funding from the Juan de la Cierva fellowship (JCI 2012-14155), the European Commission (STREP `RAQUEL') and the Spanish MINECO Project No. FIS2013-40627-P, the Generalitat de Catalunya CIRIT Project No. 2014 SGR 966 and fruitful discussions at LIQUID. This project was supported by was supported by the Austrian Academy of Sciences (ÖAW), the European Research Council (SIQS Grant No. 600645 EU-FP7-ICT), the Austrian Science Fund (FWF) with SFB F40 (FOQUS). 

\renewcommand\thefigure{\arabic{figure}}
\renewcommand{\figurename}{Extended Data Figure}        

\section*{Methods}

\setcounter{figure}{0}

\textbf{Down-conversion sources.} As shown in extended data Fig.~1, a femtosecond pulsed Ti:Sapphire laser (Coherent Chameleon Ultra II) with a center wavelength of 808nm is focused by lens L1 into a $\beta$-Barium Borate crystal (BBO) to generate blue pump pulses at 404nm via the process of second-harmonic generation. The BBO crystal is incrementally moved by a motor (M) in order to avoid damage to it. The generated pump pulses (average power of 820mW) are re-collimated by lens L2 and separated from the 808nm light by a dichroic mirror (DM1). They are then focused with lens L3 onto two periodically poled potassium titanyl phosphate (ppKTP) crystals (NL1 and NL2) with dimension $1\text{mm}\times2\text{mm}\times1\text{mm}$ and poling period $9.55\mu\text{m}$ for type-II phase matching. Two independent photon pairs entangled in their orbital angular momentum (OAM) are generated in the two crystals via spontaneous parametric down-conversion of the same pump pulse. The crystals are spatially oriented so that down-conversion occurs when the pump pulses are vertically polarized. In order to conform to the phase-matching conditions for producing wavelength-degenerate photon pairs at 808nm, NL1 (NL2) is heated to 119.1$^\circ$C (121.9$^\circ$C). A $4f$ imaging system composed of two lenses (L4) between between the two crystals ensures that the waist of the pump pulse is approximately 240$\mu$m at both crystals. The pump waist is chosen such that generation rate of $l=0$ photons is approximately twice that of the $l=\pm1$ photons in the two OAM-entangled photon pairs. Note that the OAM state distributions in equation \eqref{tensor9} are flat, while our experimentally generated distributions are not. We account for this by unbalancing the trigger photon (D) state superposition to $\ket{P}^{0,\text{-}1}_D=0.51\ket{0}_D+0.86\ket{\text{-}1}_D$, which surprisingly lowers our average four-fold count rates only by 6\%. Two dichroic mirrors (DM2) separate the 404nm pump pulses from the entangled photon pairs. Due to the high pump power, the Kerr-lensing effect cannot be neglected and adds an equivalent lens with 1m focal length into the optical path. This is compensated by carefully aligning the $4f$ imaging system (L4). To avoid gray-tracking due to the high pump power, the ppKTP crystals are also moved incrementally in steps of 200$\mu$m with motors (M). Movement of the BBO and ppKTP crystals shows slight variation in the count rates, which is compensated for by averaging over several periods of movement within each measurement interval. Two polarizing beam splitters (PBS) deterministically separate both photon pairs such that all four photons are propagating in different directions.\\

\begin{figure}[h]
\center
  \includegraphics[width=1\textwidth]{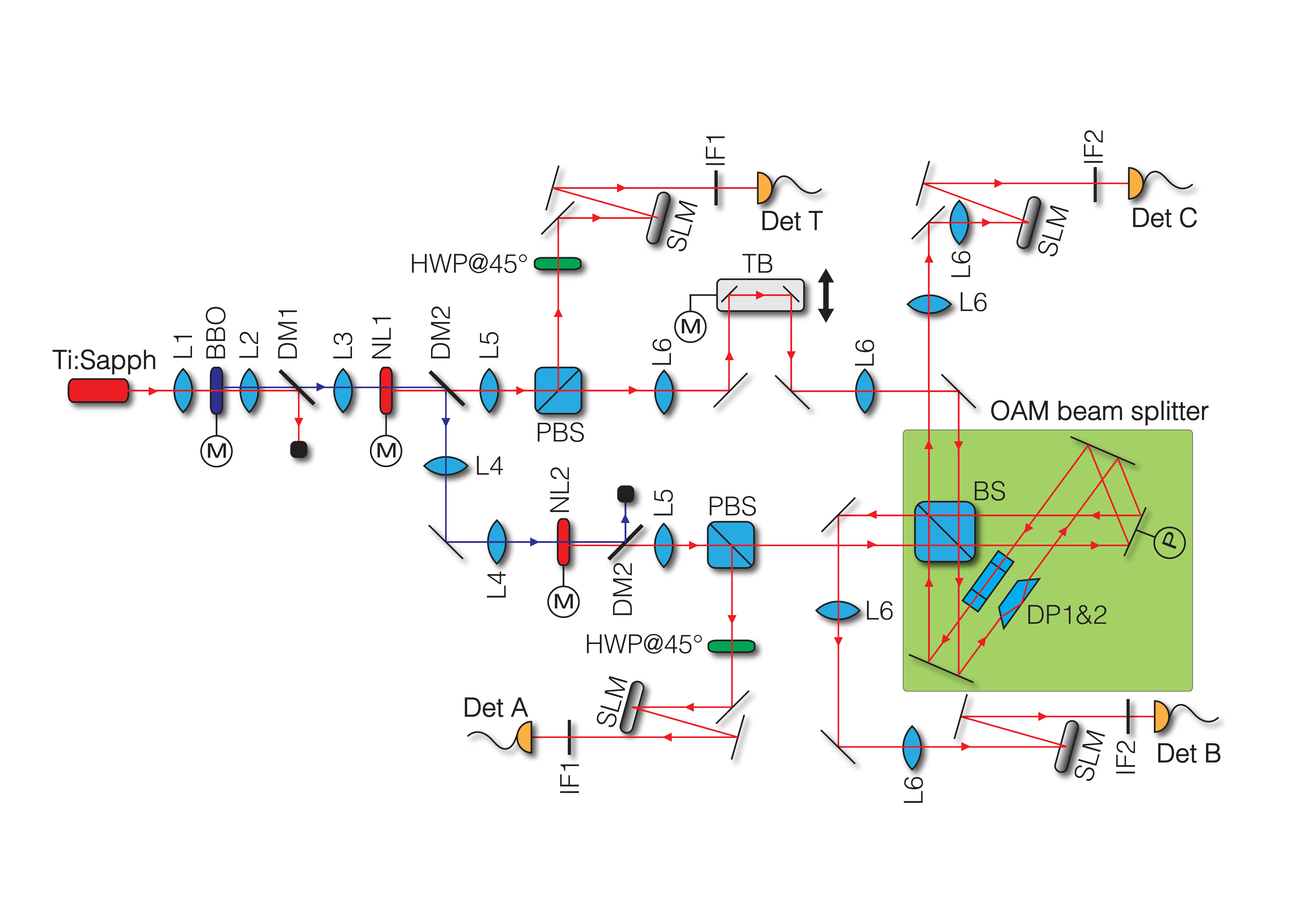}
	\caption{\label{fig:332_supp-sketch-detailed}\textbf{Detailed Experimental Setup.} A Ti:Sapphire pulsed laser source with pulse duration of 140fs and repetition rate of 80 MHz centered at 808nm is focused by lens L1 into a $\beta$-Barium Borate crystal (BBO) to produce pump pulses at 404nm via the process of second harmonic generation. In order to avoid damage, the BBO crystal is moving periodically with a motor (M). Lens L2 re-collimates the 404nm pump pulses, which are separated from the 808nm laser by a dichroic mirror (DM1). Two OAM-entangled photon pairs are produced via type-II spontaneous parametric down-conversion in two periodically poled Potassium Titanyl Phosphate (ppKTP) crystals (NL1 and NL2) of dimensions $1\times 2\times 1$mm. The full-width half-maximum beam waist of the 404nm pump laser at the crystals is 240$\mu$m with an average power of 820mW. To prevent gray-tracking and crystal damage, the ppKTP crystals are moved periodically up and down in steps of 200$\mu$m via motors (M). A second dichroic mirror (DM2) separates the entangled photon pairs from the 404nm pump laser beam. Lenses L5 are used to collimate the down-converted 808nm photon pairs. A $4f$ imaging system L4 is used to perfectly image the 404nm pump beam from crystal NL1 to crystal NL2 to ensure that the pump beam mode is exactly the same at both crystals. Polarizing beam splitter (PBS) are used to deterministically separate all four photons. A moving trombone system of mirrors (TB) is used to ensure that photons from both crystals arrive at the OAM beam splitter at the same time. A $4f$ system of lenses (L6) compensates the extra propagation that the photon from crystal NL1 undergoes, ensuring that both photons have a good spatial mode overlap at the entrance to the OAM beam splitter (highlighted in green). The OAM beam splitter is implemented in a double Sagnac interferometer configuration that allows for stable operation over several days. Note that this is different from the physical implementation shown in Fig.~\ref{setup} (for simplicity), but has exactly the same physical outcome. A piezo controlled mirror (P) is scanned every hour to optimize the alignment of the interferometer and ensure that it is sorting OAM modes correctly. Two $4f$ systems of lenses (L6) image the output of the OAM beam splitter onto spatial light modulators (SLMs). Two half-wave plates (HWPs) at 45$^\circ$ rotate the photon polarization from vertical to horizontal in paths A and D. Projective measurements are performed  by four SLMs in combination with aspheric lenses and single mode fibers, which guide the photons to single photon avalanche diodes (Det A, B, C, and T). Additionally, narrowband filters (IF1 and IF2) are used before all four detectors to ensure spectral indistinguishability between the interfering photons.}
\end{figure}

\noindent\textbf{OAM beam splitter.} This is an interferometric two-input two-output device that mixes input photons based on the parity of their OAM mode. In Fig.~\ref{setup} in the text, the OAM beam splitter is depicted for simplicity in its original design\cite{Leach:2002wy} as a Mach-Zehnder interferometer. However, our experiment requires long-term interferometric stability, which is not easily attainable with a Mach-Zehnder interferometer. For this reason, we implement the OAM beam splitter in a zero-area, double Sagnac configuration. Highlighted in green in extended data Fig.~\ref{fig:332_supp-sketch-detailed}, this configuration takes a standard Sagnac loop (which is one-input and one-output) and shifts it laterally, creating two counter-propagating Sagnac loops side-by-side that meet back up at the beam splitter. This allows us to install a dove prism (DP1 and DP2) in each loop. The interferometer is further implemented in a zero-area configuration that is rotationally invariant, minimizing the rotational dependence of the entire interferometer on the rotation of the earth. Furthermore, a piezo controlled mirror (P) is scanned every hour to optimize the alignment of the interferometer and ensure it is sorting OAM modes correctly. In addition to controlling one of the mirrors actively, the interferometer is enclosed in a plastic box that reduces the air flow through the device and stabilizes the temperature to within $\pm$0.02$\degree$C.
All of these efforts lead to a very long stability ($>80$ hours) and at the same time achieve a very high sorting efficiency of even and odd OAM modes (99:1).\\

\noindent\textbf{Interference of independent photons.}\label{chap_supp:four-photon-HOM-effect} The interference of two photons at the OAM beam splitter crucially depends on the indistinguishability of the two particles involved. Since the presence of the interfering photons is indicated by the trigger detection events of their respective partner photons, we effectively observe four-photon coincidence events. The interference visibility strongly depends on the physical properties of these four photons, as well as the process by which they are created. This particular phenomenon has been studied in great detail\cite{Kaltenbaek:2009vf} and the theoretically expected visibility is calculated to have the form
\begin{align}
V'=\Bigg[ 2 \frac{\sqrt{\sigma_T^2+\sigma_P^2}\sqrt{\sigma_S^2+\sigma_P^2+\sigma_S^2\sigma_P^2\tau_J^2}}{\sigma_P\sqrt{\sigma_S^2+\sigma_T^2+\sigma_P^2}}-1 \Bigg]^{-1}.
\end{align}
Here, $\sigma_P$, $\sigma_S$, and $\sigma_T$ are the gaussian spectral widths of the pump, signal, and trigger photons respectively (in units of frequency). $\sigma_S$ and $\sigma_T$ are determined by the widths of narrowband interference filters IF2 and IF1, respectively. $\tau_J$ has units of time and refers to the relative timing jitter between the two crystals, which arises due to the group velocity mismatch between the 404nm pump pulse and the 808nm down-converted photons in the ppKTP crystals. Additionally, due to the finite pulse width of the 404nm pump beam, a second relative timing jitter between the two crystals is also present. In our experiment, the sum of these two timing jitters is approximately 1ps~\cite{scheidl}.

In addition to spectral distinguishability, the spatial and temporal distinguishability of the two photons both play an important role in the interference visibility. The moving trombone system of mirrors (TB) is used to find the position of zero-delay when photons from both crystals arrive at the OAM beam splitter at the same exact time. The spatial-mode distinguishability of the two input photons at the OAM beam splitter has a further detrimental effect on the total interferometric visibility. We use a $4f$ imaging system of lenses (L6) in order to compensate for the extra propagation that the photon from NL1 undergoes with respect to the photon from NL2. Nonetheless, the mode overlap of the two photons at the OAM splitter is not perfect, and modifies the two-photon interference visibility in the following way:

\begin{align}
V=\frac{\eta^2 V'}{1+V'(1-\eta^2)}.
\end{align}
Here, $\eta=\eta_{\text{OAM}}\times \eta_{\text{SP}}$ is the product of the sorting efficiency $\eta_{\text{OAM}}$ of the OAM beam splitter and the spatial mode overlap $\eta_{\text{SP}}$ of the two input photons. Inserting our measured/estimated experimental parameters ($\sigma_P=3.67\text{THz}$, $\sigma_T=588\text{GHz}$, $\sigma_S=184\text{GHz}$, $\tau_J=1\text{p}s$, $\eta_{\text{OAM}}=0.99$ and $\eta_{\text{SP}} =0.9$) into the above formula yields a two-photon interference visibility of $V=0.64$, which is very close to the experimental visibility observed in Fig.~\ref{HOM}. The visibility crucially depends on the spectral width $\sigma_S$ of the two interfering photons. A smaller $\sigma_S$ greatly improves the interference visibility at the cost of lowering the four-photon count rates. It is therefore important to find an optimal value that provides both, sufficiently high visibility and practical count rates. In our experiment, we found this value to be $\sigma_S=184\text{GHz}$. Assuming a Gaussian spectral distribution, the theoretical full-width at half-maximum (FWHM) of the two-photon interference dip is then given by

\be L_{\text{dip}}=\frac{c}{\pi}\sqrt{\frac{2\text{ln}2}{\sigma_P^{-2}+\sigma_S^{-2}+\tau_J^2}}.\ee

\noindent Inserting the corresponding values from above leads to a expected FWHM dip-width of 624$\mu$m, which is close to our experimentally observed value of 473$\mu$m in Fig.~\ref{HOM}.\\

\noindent\textbf{Single Photon Projective Measurements.} Two additional $4f$ imaging systems (L6) image the output of the OAM beam splitter onto two holographic spatial light modulators (SLMs). These are liquid-crystal devices that can impart an arbitrary 2D phase pattern onto a photon. We use these devices to perform projection measurements of OAM. These measurements are carried out by flattening the phase of incident photons such that they couple into a single-mode optical fiber\cite{mair}. The scheme is based on the fact that a single-mode fiber (SMF) only carries spatial modes with $\ell=0$. In order to detect a photon with $\ell=1$, a hologram with $\ell=-1$ displayed on the SLM projects the photon into mode $\ell=0$. The photon then couples efficiently to an SMF and is guided to an single photon avalanche photodiode (Excelitas SPCM-AQRH-14-FC). The same procedure also applies to projective measurements of mode superpositions. Note that in our experiment, phase-only holograms are used, which take only the vortex phase structure $e^{i\ell\phi}$ of a Laguerre-Gaussian mode into account, and not its amplitude. This results in an imperfect mode overlap at the SMF~\cite{quassim} and leads to lowered coupling efficiencies for modes with $\ell=\pm 1$. This drop in efficiency can be somewhat compensated by adding a mode-dependent quadratic phase to these holograms, which changes the effective coupling mode-waist.\\

\noindent\textbf{$\mathbf{(332)}$-Entanglement witness.} In order to prove that the state is indeed a $(332)$-type entangled state we have to prove that it cannot be decomposed into states of a smaller dimensionality structure. We thus have to show that it lies outside the $(322)$-set of states, that is the convex hull of all states that can be decomposed into $322$ and $232$ states. From the measured data we can extract the fidelity to the ideal state
\begin{equation}
	\ket{\Psi}=\frac{1}{\sqrt{3}}\left( \ket{0,0,0}+\ket{1,-1,1}+\ket{-1,1,1} \right)\,,
	\label{ideal}
\end{equation}
which we will denote as $F_\text{exp}:=\text{Tr}(\rho_\text{exp}|\Psi\rangle\langle\Psi|)$. We thus need to compare the experimental fidelity with the best achievable fidelity of a $(322)$-state, i.e.
\begin{align}
F_\text{max}:=\max_{\sigma\in (322)}\text{Tr}(\sigma |\Psi\rangle\langle\Psi|)\,.
\end{align}
If $F_\text{exp}>F_\text{max}$, we can conclude that the experimentally certified fidelity cannot be explained by any state in $(322)$ and thus the underlying state is certified to have an entangled dimensionality structure of $(332)$. To calculate $F_\text{max}$ it is useful to observe that it is convex in the set of states, i.e. the maximum will always also be reachable by a pure state. Furthermore, since the set $(322)$ is the convex hull of $322$ and $232$, we can write the following
\begin{align}
\label{first}
F_\text{max}=\max_{|\Phi\rangle\in (322)}|\langle\Psi|\Phi\rangle|^2=\text{max}[\max_{|\Phi\rangle\in 322}|\langle\Psi|\Phi\rangle|^2,\max_{|\Phi\rangle\in 232}|\langle\Psi|\Phi\rangle|^2]\,.
\end{align}


\noindent Now for a fixed rank vector $xyz$, these fidelities can be bounded by noticing that
\begin{align}
\label{second}
\max_{|\Phi\rangle\in xyz}|\langle\Psi|\Phi\rangle|^2\leq\min[\max_{rank(\text{Tr}_{23}|\Phi\rangle\langle\Phi|)=x}|\langle\Psi|\Phi\rangle|^2,\max_{rank(\text{Tr}_{13}|\Phi\rangle\langle\Phi|)=y}|\langle\Psi|\Phi\rangle|^2,\max_{rank(\text{Tr}_{12}|\Phi\rangle\langle\Phi|)=z}|\langle\Psi|\Phi\rangle|^2]\,.
\end{align}
Now inserting (\ref{second}) in (\ref{first}) we can compute each of the appearing terms using a previously published theorem~\cite{Fickler:2014eq}. It states that if a state has a Schmidt decomposition across a cut $A|\overline{A}$ given by $|\Psi\rangle=\sum_{i_0}^{r-1}\lambda_i|v_{A}^i\rangle\otimes|v_{\overline{A}}^i\rangle$, we can compute the maximal overlap with a state of bounded rank across this partition as
\begin{align}
\max_{rank(\text{Tr}_{\overline{A}}|\Phi\rangle\langle\Phi|)=x}|\langle\Psi|\Phi\rangle|^2=\sum_{i=0}^{x-1}\lambda_i^2\,,
\end{align}
where we assumed ordered Schmidt coefficients, i.e. $\lambda_i\geq\lambda_{i+1}$. Now all we need are the coefficients for the Schmidt decompositions for our target state for all three partitions. For $3|12$ they are $\{\frac{\sqrt{2}}{\sqrt{3}},\frac{1}{\sqrt{3}}\}$ and for $2|13$ and $1|23$ we get $\{\frac{1}{\sqrt{3}},\frac{1}{\sqrt{3}},\frac{1}{\sqrt{3}}\}$. Inserting these numbers we find that the maximum overlap of the target state (\ref{ideal}) with a $(322)$ state is given by 
\begin{align}
F_\text{max}=\frac{2}{3}\,.
\end{align}

\noindent\textbf{Witness Measurements.} The experimental fidelity $F_{\text{exp}}:=\text{Tr}(\rho_{\text{exp}}|\Psi\rangle\langle\Psi|)$ determines which measurements are required. The projector $|\Psi\rangle\langle\Psi|$ projects only onto the non-zero diagonal and off-diagonal elements contained in the density matrix $\rho_{\text{exp}}$. Additionally, for the purpose of normalization, it is necessary to measure all other diagonal elements in $\rho_{\text{exp}}$. This results in 18 diagonal and 3 unique off-diagonal elements that need to be measured in order to calculate $F_{\text{exp}}$. In our experiment, we can only perform projective measurements with SLMs. A diagonal element is given by one single projection $\langle ijk|\rho|ijk\rangle=\frac{C(ijk)}{C_T}$, with $C_T:=\sum_{i=-1,0,1}\sum_{j=-1,0,1}\sum_{k=0,1}C(ijk)$ containing all diagonal elements for normalization. Out of the six off-diagonal elements, only three are unique and need to be measured: $\langle 000|\rho|1\text{-}11\rangle$, $\langle 000|\rho|\text{-}111\rangle$ and $\langle \text{-}111|\rho|1\text{-}11\rangle$. Note that the last off-diagonal element is only in a two-particle superposition. Hence, it can be measured in the standard way that two-particle two-dimensional states are usually measured. In order to measure the other two off-diagonal elements with projective measurements, we decompose them into $\sigma_\text{x}$ and $\sigma_\text{y}$ measurements. The real and imaginary part of each element can be written as 
\begin{eqnarray}\label{eq:real-imaginary-off-diagonals}
\Re\big[ \langle ijk|\rho|lmn\rangle \big]=\expval{\sig{x}{i}{l}\otimes\sig{x}{j}{m}\otimes\sig{x}{k}{n}}-\expval{\sig{y}{i}{l}\otimes\sig{y}{j}{m}\otimes\sig{x}{k}{n}}-\expval{\sig{y}{i}{l}\otimes\sig{x}{j}{m}\otimes\sig{y}{k}{n}}-\expval{\sig{x}{i}{l}\otimes\sig{y}{j}{m}\otimes\sig{y}{k}{n}}\\\nonumber
\Im\big[ \langle ijk|\rho|lmn\rangle \big]=\expval{\sig{y}{i}{l}\otimes\sig{y}{j}{m}\otimes\sig{y}{k}{n}}-\expval{\sig{x}{i}{l}\otimes\sig{x}{j}{m}\otimes\sig{y}{k}{n}}-\expval{\sig{x}{i}{l}\otimes\sig{y}{j}{m}\otimes\sig{x}{k}{n}}-\expval{\sig{y}{i}{l}\otimes\sig{x}{j}{m}\otimes\sig{x}{k}{n}},
\end{eqnarray}
where $ \sig{x}{a}{b}=\ket{a}\bra{b}+\ket{b}\bra{a} $ and $ \sig{y}{a}{b}=i \ket{a}\bra{b} - i \ket{b}\bra{a} $. The $\sigma_{x,y}$ operators are also not measurable directly with SLMs and are therefore rewritten using the following operators:
\begin{eqnarray}
	\widehat{\mathcal{P}}_+(a,b)=\ket{+}\bra{+}_{(a,b)}=\ket{a}\bra{a}+\ket{b}\bra{b}+\ket{a}\bra{b}+\ket{b}\bra{a}\\\nonumber
	\widehat{\mathcal{P}}_-(a,b)=\ket{-}\bra{-}_{(a,b)}=\ket{a}\bra{a}+\ket{b}\bra{b}-\ket{a}\bra{b}-\ket{b}\bra{a}\\\nonumber
	\widehat{\mathcal{P}}_{+i}(a,b)=\ket{+i}\bra{+i}_{(a,b)}=\ket{a}\bra{a}+\ket{b}\bra{b}-i\ket{a}\bra{b}+i\ket{b}\bra{a}\\\nonumber
	\widehat{\mathcal{P}}_{-i}(a,b)=\ket{-i}\bra{-i}_{(a,b)}=\ket{a}\bra{a}+\ket{b}\bra{b}+i\ket{a}\bra{b}-i\ket{b}\bra{a},
\end{eqnarray}
where $\ket{+}_{(a,b)}=\ket{a}+\ket{b}$, $\ket{-}_{(a,b)}=\ket{a}-\ket{b}$, $\ket{+i}_{(a,b)}=\ket{a}+i\ket{b}$ and $\ket{-i}_{(a,b)}=\ket{a}-i\ket{b}$. These superposition states can be measured with SLMs in our experiment. Thus the $\sigma$ operators from Eq.~\ref{eq:real-imaginary-off-diagonals} can be written in the following manner:
\begin{eqnarray}
	\sig{x}{a}{b}=\frac{1}{2}\left( \widehat{\mathcal{P}}_+(a,b) -\widehat{\mathcal{P}}_-(a,b) \right)\nonumber\\
	\sig{y}{a}{b}=\frac{1}{2}\left( \widehat{\mathcal{P}}_{-i}(a,b) -\widehat{\mathcal{P}}_{+i}(a,b) \right).
\end{eqnarray}
This leads to 64 projection measurements required for measuring each of the two off-diagonal elements $\langle 000|\rho|1\text{-}11\rangle$ and $\langle 000|\rho|\text{-}111\rangle$. Only 16 projection measurements are required for the off-diagonal element  $\langle \text{-}111|\rho|1\text{-}11\rangle$ since it involves only one dimension for photon $C$. Summing up these measurements as well as the 18 diagonal elements leads to 162 total measurements. The results of these are given in extended data table 1. From these measurements, the overlap between the generated state $\rho_\text{exp}$ and the ideal state $(332)$-state $\ket{\Psi}$ is calculated to be $F_\text{exp}=0.801\pm0.018$. The error in the overlap is calculated by propagating the Poissonian error in the photon-counting rates by performing a Monte Carlo simulation of the experiment.

\let\oldthebibliography=\thebibliography
\let\oldendthebibliography=\endthebibliography
\renewenvironment{thebibliography}[1]{%
    \oldthebibliography{#1}%
    \setcounter{enumiv}{ 30 }%
}{\oldendthebibliography}

\renewcommand\thefigure{\arabic{figure}}
\renewcommand{\figurename}{Extended Data Table}        

\setcounter{figure}{0}

\begin{figure}[h]
\center
  \includegraphics[width=1\textwidth]{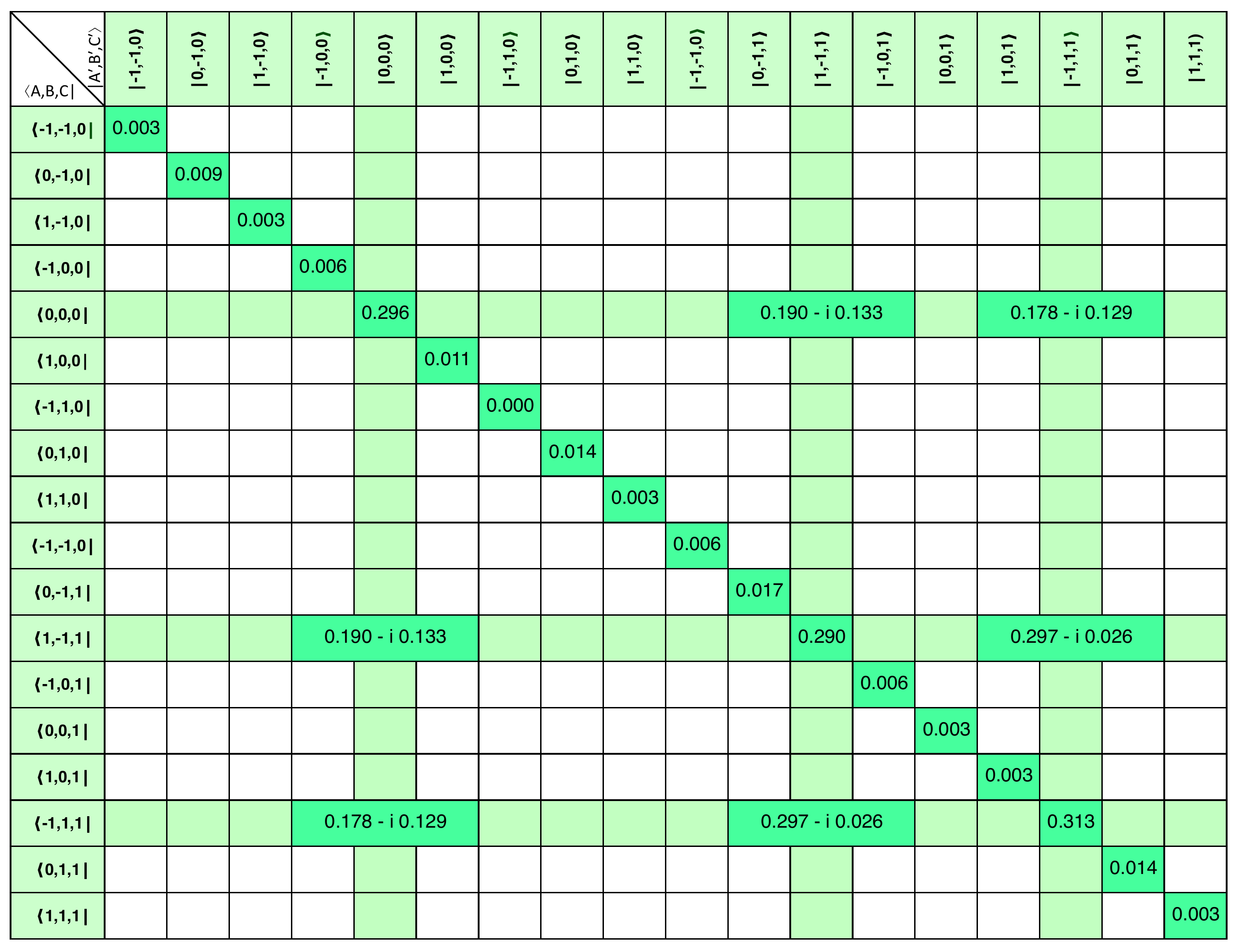}
  \caption{Numerical values of the measured density matrix elements.}
\end{figure}


\begin{thebibliography}{10}
\expandafter\ifx\csname url\endcsname\relax
  \def\url#1{\texttt{#1}}\fi
\expandafter\ifx\csname urlprefix\endcsname\relax\def\urlprefix{URL }\fi
\providecommand{\bibinfo}[2]{#2}
\providecommand{\eprint}[2][]{\url{#2}}

\bibitem{Einstein:1935hx}
\bibinfo{author}{Einstein, A.}, \bibinfo{author}{Podolsky, B.} \&
  \bibinfo{author}{Rosen, N.}
\newblock \bibinfo{title}{{Can Quantum-Mechanical Description of Physical
  Reality Be Considered Complete?}}
\newblock \emph{\bibinfo{journal}{Phys. Rev.}} \textbf{\bibinfo{volume}{47}},
  \bibinfo{pages}{777--780} (\bibinfo{year}{1935}).

\bibitem{Bell:1964wu}
\bibinfo{author}{Bell, J.}
\newblock \bibinfo{title}{{On the einstein-podolsky-rosen paradox}}.
\newblock \emph{\bibinfo{journal}{Physics}} \textbf{\bibinfo{volume}{1}},
  \bibinfo{pages}{195--200} (\bibinfo{year}{1964}).

\bibitem{Bennett:2000kl}
\bibinfo{author}{Bennett, C.~H.} \& \bibinfo{author}{DiVincenzo, D.~P.}
\newblock \bibinfo{title}{{Quantum information and computation}}.
\newblock \emph{\bibinfo{journal}{Nature}} \textbf{\bibinfo{volume}{404}},
  \bibinfo{pages}{247--255} (\bibinfo{year}{2000}).

\bibitem{Ekert:1991kl}
\bibinfo{author}{Ekert, A.~K.}
\newblock \bibinfo{title}{{Quantum cryptography based on Bell{\textquoteright}s
  theorem}}.
\newblock \emph{\bibinfo{journal}{Phys. Rev. Lett.}}
  \textbf{\bibinfo{volume}{67}}, \bibinfo{pages}{661--663}
  (\bibinfo{year}{1991}).

\bibitem{Greenberger:1989vx}
\bibinfo{author}{Greenberger, D.~M.}, \bibinfo{author}{Horne, M.~A.} \&
  \bibinfo{author}{Zeilinger, A.}
\newblock \bibinfo{title}{{Going Beyond Bell's Theorem}}.
\newblock In \bibinfo{editor}{Kafatos, M.} (ed.)
  \emph{\bibinfo{booktitle}{Bell's Theorem, Quantum Theory, and Conceptions of
  the Universe}}, \bibinfo{pages}{69--72} (\bibinfo{address}{Kluwer,
  Dordrecht}, \bibinfo{year}{1989}).

\bibitem{Pan:2000do}
\bibinfo{author}{Pan, J.-W.}, \bibinfo{author}{Bouwmeester, D.},
  \bibinfo{author}{Daniell, M.}, \bibinfo{author}{Weinfurter, H.} \&
  \bibinfo{author}{Zeilinger, A.}
\newblock \bibinfo{title}{{Experimental test of quantum nonlocality in
  three-photon Greenberger{\textendash}Horne{\textendash}Zeilinger
  entanglement}}.
\newblock \emph{\bibinfo{journal}{Nature}} \textbf{\bibinfo{volume}{403}},
  \bibinfo{pages}{515--519} (\bibinfo{year}{2000}).

\bibitem{Kelly:2015gi}
\bibinfo{author}{Kelly, J.} \emph{et~al.}
\newblock \bibinfo{title}{{State preservation by repetitive error detection in
  a superconducting quantum circuit}}.
\newblock \emph{\bibinfo{journal}{Nature}} \textbf{\bibinfo{volume}{519}},
  \bibinfo{pages}{66--69} (\bibinfo{year}{2015}).

\bibitem{Yao:2012fp}
\bibinfo{author}{Yao, X.-C.} \emph{et~al.}
\newblock \bibinfo{title}{{Observation of eight-photon entanglement}}.
\newblock \emph{\bibinfo{journal}{Nat. Phot.}} \textbf{\bibinfo{volume}{6}},
  \bibinfo{pages}{225--228} (\bibinfo{year}{2012}).

\bibitem{Lanyon:2014eh}
\bibinfo{author}{Lanyon, B.~P.} \emph{et~al.}
\newblock \bibinfo{title}{{Experimental Violation of Multipartite Bell
  Inequalities with Trapped Ions}}.
\newblock \emph{\bibinfo{journal}{Phys. Rev. Lett.}}
  \textbf{\bibinfo{volume}{112}}, \bibinfo{pages}{100403}
  (\bibinfo{year}{2014}).

\bibitem{Huber:2013ie}
\bibinfo{author}{Huber, M.} \& \bibinfo{author}{de~Vicente, J.}
\newblock \bibinfo{title}{{Structure of Multidimensional Entanglement in
  Multipartite Systems}}.
\newblock \emph{\bibinfo{journal}{Phys. Rev. Lett.}}
  \textbf{\bibinfo{volume}{110}}, \bibinfo{pages}{030501}
  (\bibinfo{year}{2013}).

\bibitem{Zeilinger:1997eb}
\bibinfo{author}{Zeilinger, A.}, \bibinfo{author}{Horne, M.},
  \bibinfo{author}{Weinfurter, H.} \& \bibinfo{author}{{\.{Z}}ukowski, M.}
\newblock \bibinfo{title}{{Three-Particle Entanglements from Two Entangled
  Pairs}}.
\newblock \emph{\bibinfo{journal}{Phys. Rev. Lett.}}
  \textbf{\bibinfo{volume}{78}}, \bibinfo{pages}{3031--3034}
  (\bibinfo{year}{1997}).

\bibitem{Hillery:1999cb}
\bibinfo{author}{Hillery, M.}, \bibinfo{author}{Bu{\v z}ek, V.} \&
  \bibinfo{author}{Berthiaume, A.}
\newblock \bibinfo{title}{{Quantum secret sharing}}.
\newblock \emph{\bibinfo{journal}{Phys. Rev. A}} \textbf{\bibinfo{volume}{59}},
  \bibinfo{pages}{1829--1834} (\bibinfo{year}{1999}).

\bibitem{Clauser:1969ff}
\bibinfo{author}{Clauser, J.}, \bibinfo{author}{Horne, M.},
  \bibinfo{author}{Shimony, A.} \& \bibinfo{author}{Holt, R.}
\newblock \bibinfo{title}{{Proposed Experiment to Test Local Hidden-Variable
  Theories}}.
\newblock \emph{\bibinfo{journal}{Phys. Rev. Lett.}}
  \textbf{\bibinfo{volume}{23}}, \bibinfo{pages}{880--884}
  (\bibinfo{year}{1969}).

\bibitem{Lapkiewicz:2011iq}
\bibinfo{author}{Lapkiewicz, R.} \emph{et~al.}
\newblock \bibinfo{title}{{Experimental non-classicality of an indivisible
  quantum system}}.
\newblock \emph{\bibinfo{journal}{Nature}} \textbf{\bibinfo{volume}{474}},
  \bibinfo{pages}{490--493} (\bibinfo{year}{2011}).

\bibitem{Mirhosseini:2015fy}
\bibinfo{author}{Mirhosseini, M.} \emph{et~al.}
\newblock \bibinfo{title}{{High-dimensional quantum cryptography with twisted
  light}}.
\newblock \emph{\bibinfo{journal}{New J. Phys.}} \textbf{\bibinfo{volume}{17}},
  \bibinfo{pages}{033033} (\bibinfo{year}{2015}).

\bibitem{Malik:2014ht}
\bibinfo{author}{Malik, M.} \& \bibinfo{author}{Boyd, R.~W.}
\newblock \bibinfo{title}{{Quantum Imaging Technologies}}.
\newblock \emph{\bibinfo{journal}{Riv Nuovo Cimento}}
  \textbf{\bibinfo{volume}{37}}, \bibinfo{pages}{273} (\bibinfo{year}{2014}).

\bibitem{MolinaTerriza:2007ig}
\bibinfo{author}{Molina-Terriza, G.}, \bibinfo{author}{Torres, J.~P.} \&
  \bibinfo{author}{Torner, L.}
\newblock \bibinfo{title}{{Twisted photons}}.
\newblock \emph{\bibinfo{journal}{Nature Physics}}
  \textbf{\bibinfo{volume}{3}}, \bibinfo{pages}{305--310}
  (\bibinfo{year}{2007}).

\bibitem{Dada:2011dn}
\bibinfo{author}{Dada, A.~C.}, \bibinfo{author}{Leach, J.},
  \bibinfo{author}{Buller, G.~S.}, \bibinfo{author}{Padgett, M.~J.} \&
  \bibinfo{author}{Andersson, E.}
\newblock \bibinfo{title}{{Experimental high-dimensional two-photon
  entanglement and violations of generalized Bell inequalities}}.
\newblock \emph{\bibinfo{journal}{Nature Physics}}
  \textbf{\bibinfo{volume}{7}}, \bibinfo{pages}{677--680}
  (\bibinfo{year}{2011}).

\bibitem{Krenn:2014jy}
\bibinfo{author}{Krenn, M.} \emph{et~al.}
\newblock \bibinfo{title}{{Generation and confirmation of a (100 x
  100)-dimensional entangled quantum system}}.
\newblock \emph{\bibinfo{journal}{PNAS}} \textbf{\bibinfo{volume}{111}},
  \bibinfo{pages}{6243} (\bibinfo{year}{2014}).

\bibitem{Terhal:2000gd}
\bibinfo{author}{Terhal, B.~M.} \& \bibinfo{author}{Horodecki, P.}
\newblock \bibinfo{title}{{Schmidt number for density matrices}}.
\newblock \emph{\bibinfo{journal}{Phys. Rev. A}} \textbf{\bibinfo{volume}{61}},
  \bibinfo{pages}{040301} (\bibinfo{year}{2000}).

\bibitem{Cadney:2014iw}
\bibinfo{author}{Cadney, J.}, \bibinfo{author}{Huber, M.},
  \bibinfo{author}{Linden, N.} \& \bibinfo{author}{Winter, A.}
\newblock \bibinfo{title}{{Inequalities for the ranks of multipartite quantum
  states}}.
\newblock \emph{\bibinfo{journal}{Linear Algebra and its Applications}}
  \textbf{\bibinfo{volume}{452}}, \bibinfo{pages}{153--171}
  (\bibinfo{year}{2014}).

\bibitem{Bouwmeester:1999jq}
\bibinfo{author}{Bouwmeester, D.}, \bibinfo{author}{Pan, J.-W.},
  \bibinfo{author}{Daniell, M.}, \bibinfo{author}{Weinfurter, H.} \&
  \bibinfo{author}{Zeilinger, A.}
\newblock \bibinfo{title}{{Observation of Three-Photon
  Greenberger-Horne-Zeilinger Entanglement}}.
\newblock \emph{\bibinfo{journal}{Phys. Rev. Lett.}}
  \textbf{\bibinfo{volume}{82}}, \bibinfo{pages}{1345--1349}
  (\bibinfo{year}{1999}).

\bibitem{Pan:2001vk}
\bibinfo{author}{Pan, J.-W.}, \bibinfo{author}{Daniell, M.},
  \bibinfo{author}{Gasparoni, S.}, \bibinfo{author}{Weihs, G.} \&
  \bibinfo{author}{Zeilinger, A.}
\newblock \bibinfo{title}{{Experimental Demonstration of Four-Photon
  Entanglement and High-Fidelity Teleportation}}.
\newblock \emph{\bibinfo{journal}{Phys. Rev. Lett.}}
  \textbf{\bibinfo{volume}{86}}, \bibinfo{pages}{4435--4438}
  (\bibinfo{year}{2001}).

\bibitem{Leach:2002wy}
\bibinfo{author}{Leach, J.}, \bibinfo{author}{Padgett, M.~J.},
  \bibinfo{author}{Barnett, S.~M.} \& \bibinfo{author}{Franke-Arnold, S.}
\newblock \bibinfo{title}{{Measuring the orbital angular momentum of a single
  photon}}.
\newblock \emph{\bibinfo{journal}{Phys. Rev. Lett.}}
  \textbf{\bibinfo{volume}{88}}, \bibinfo{pages}{257901}
  (\bibinfo{year}{2002}).

\bibitem{Hong:1987gm}
\bibinfo{author}{Hong, C.}, \bibinfo{author}{Ou, Z.} \&
  \bibinfo{author}{Mandel, L.}
\newblock \bibinfo{title}{{Measurement of subpicosecond time intervals between
  two photons by interference}}.
\newblock \emph{\bibinfo{journal}{Phys. Rev. Lett.}}
  \textbf{\bibinfo{volume}{59}}, \bibinfo{pages}{2044--2046}
  (\bibinfo{year}{1987}).

\bibitem{Kaltenbaek:2009vf}
\bibinfo{author}{Kaltenbaek, R.}
\newblock \emph{\bibinfo{title}{{Interference of Photons from Independent
  Sources}}}.
\newblock Ph.D. thesis, \bibinfo{school}{University of Vienna}
  (\bibinfo{year}{2009}).

\bibitem{Fickler:2014eq}
\bibinfo{author}{Fickler, R.} \emph{et~al.}
\newblock \bibinfo{title}{{Interface between path and orbital angular momentum
  entanglement for high-dimensional photonic quantum information}}.
\newblock \emph{\bibinfo{journal}{Nat. Commun.}}
  \textbf{\bibinfo{volume}{5:4502}} (\bibinfo{year}{2014}).

\bibitem{Mirhosseini:2013em}
\bibinfo{author}{Mirhosseini, M.}, \bibinfo{author}{Malik, M.},
  \bibinfo{author}{Shi, Z.} \& \bibinfo{author}{Boyd, R.~W.}
\newblock \bibinfo{title}{{Efficient separation of the orbital angular momentum
  eigenstates of light}}.
\newblock \emph{\bibinfo{journal}{Nat. Commun.}}
  \textbf{\bibinfo{volume}{4:2781}} (\bibinfo{year}{2013}).

\bibitem{Melvin}
\bibinfo{author}{Krenn, M.}, \bibinfo{author}{Malik, M.},
  \bibinfo{author}{Fickler, R.}, \bibinfo{author}{Lapkiewicz, R.} \&
  \bibinfo{author}{Zeilinger, A.}
\newblock \bibinfo{title}{{Automated search for new quantum experiments}}.
\newblock \emph{\bibinfo{journal}{To appear on ArXiv}}  (\bibinfo{year}{2015}).

\bibitem{LMQKD}
\bibinfo{author}{Malik, M.} \&
  \bibinfo{author}{Huber, M.}
\newblock \bibinfo{title}{{Layered quantum cryptography}}.
\newblock \emph{\bibinfo{journal}{To appear on ArXiv}}  (\bibinfo{year}{2015}).


\end{thebibliography}

\begin{thebibliography}{99}

\bibitem{scheidl} Scheidl, Thomas, et al. ``Crossed-crystal scheme for femtosecond-pulsed entangled photon generation in periodically poled potassium titanyl phosphate." Phys. Rev. A 89, 042324 (2014). 
\bibitem{mair} Mair, Alois, et al. ``Entanglement of the orbital angular momentum states of photons." Nature 412, 313-316 (2001).
\bibitem{quassim} Qassim, Hammam, et al. ``Limitations to the determination of a Laguerre-€"Gauss spectrum via projective, phase-flattening measurement." JOSA B 31, A20-A23 (2014).
\end{thebibliography}
\end{document}